# Efficient Stratification Method for Socio - Economic Survey in Remote Areas

**Adhi Kurniawan**
Directorate of Census and Survey Methodology Development, BPS - Statistics Indonesia, Jakarta, Indonesia – adhikurnia22@gmail.com

**Atika Nashirah Hasyyati***
Directorate of Finance, IT, and Tourism Statistics, BPS - Statistics Indonesia, Jakarta, Indonesia – atikanashirah@gmail.com

**Abstract**

The problems that exist in implementing a sampling design for socio-economic surveys in remote areas in Indonesia are high cost of the survey, low response rate, and less accurate. Therefore, the sampling design needs to be developed, one of which is to improve the efficiency of the stratification procedure. Stratification of census block in remote areas can be developed by combining the strata of welfare concentration and the strata of geographic difficulty by simulating the various alternatives number of strata and the various alternatives sample allocation. The strata of welfare concentration and the strata of geographic difficulty are constructed by Polychoric Principal Component Analysis. The strata of welfare concentration aim to improve statistical efficiency, while the strata of geographic difficulty are used to improve cost efficiency. The estimation procedure is performed at the domain level and population level. The simulation study focus on Papua Province by using the 2010 Population Census data and the 2011 Village Potency data. Some sampling scenarios can be categorized into four quadrants, the first quadrant with small sampling variance and low cost, the second quadrant with big sampling variance and low cost, the third quadrant with big sampling variance and high cost, and the fourth quadrant with small sampling variance and high cost. Based on these simulation results, several alternative scenarios of efficient stratification with small sampling variance and low cost of the survey are obtained.

## 1. INTRODUCTION

Survey design has the important role in influencing the quality of data. There are two aspects that have to be considered in designing survey methods, namely statistical efficiency and the cost efficiency. Statistical efficiency means that the survey design must ensure that the estimator provided by the survey has the appropriate result regarding the reliability criteria. Practically, it is common to use the variance estimation to measure this criteria. On the other hand, cost efficiency also must be considered by the survey designer because usually cost is the commonly important constraint when conducting survey data collection. Therefore, the sample distribution have to be designed by considering two aspects of efficiency as in Kish (2004) stated that variance and cost factors are best to be viewed together.

In order to make the good statistical inference and fulfill the term of representativeness, the sample distribution must be spread out over the areas of population, including the remote areas. In this case, the remote areas issues in Indonesia must be concerned due to the heterogenous characteristics of the areas. Previous experiences in the social economic survey show that there are several problems regarding the samples existence in the remote area, such as the unpunctuality of collecting data, time schedule, accessibility problems, low response rate, and the high cost to access the areas. Therefore, it is very necessarry to develop alternative sampling method that can enhance statistical efficiency and cost efficiency in the domains which consist of the large number of remote area.

One of the most important point that can be improved in sampling design in regard with increasing statistical efficiency and cost efficiency is stratification methods. Stratification means that the population of N units is divided into subpopulations or known as strata (Cochran, 1965). The most common reason of stratification implementation in probability sampling is that stratification can increase efficiency (Rossi, Wright, and Anderson, 2013). Stratification methods can be used to decrease the variances so that there will be homogeneity within strata (Kish, 1965). Several factors that influence stratified sampling efficiency of the estimator of population parameters are the choice of stratification variable, number of strata, determination of strata boundaries, and allocation of sample sizes to the different strata (Verma, Joorel, and Agnihotri, 2012).

This paper propose alternative methods of stratification, especially for the socio-economic survey, by combining the concentration welfare stratum and geographic difficulty stratum and simulating the various alternatives the number of stratum. Stratification is constructed by classifying census block into some categories or stratum based on wealth index and geographic difficulty index. By these procedures, elements in the same stratum are expected to have the similar characteristics so that the variation within stratum will be minimum. In regard to cost efficiency, there is a possibility in stratification procedures to design the appropriate sample allocation to each stratum in order to achieve the possible sample that has the minimum cost.

## 2. GEOGRAPHIC DIFFICULTY INDEX

Data source that is used to construct geographic difficulty stratum is Village Potency 2011. This data is a village census data in Indonesia that contains various characteristics regarding the condition of the village. In order to construct the geographic difficulty stratum, there are 7 categorical variables used, such as location of village (var1), slope of land (var2), village location of forest area (var3), transportation between villages (var4), road accessibility between village (v5), distance from village to the centre of district (v6), and road accessibility of four-wheels motor vehicles within village (v7). Rescoring must be done in each variables to create the ordinal scale.

**Table 1. Scoring of Geographic Variables**

| Code | Name of variable | Score |
|---|---|---|
| var1 | Location of village | Peak of mountain |
|  |  | Mountainside |
|  |  | Valley |
|  |  | Plain |
| var2 | Slope of land | Steep |
|  |  | Medium |
|  |  | Sloping |
| var3 | Village location of forest area | Inside the forest area |
|  |  | In the edge of the forest area |
|  |  | Outside the forest area |
| var4 | Transportation between villages | Water transportation |
|  |  | Land trasportation and the road surface is ground |
|  |  | Land transportation and the road surface is hardened by gravel/stones |
|  |  | Both land transportation and water transportation, land trasportation and the road surface is asphalt |
| var5 | Road accessibility between villages | Water transportation |
|  |  | Sometimes can not be accessed |
|  |  | Always can be acceesed |
| var6 | Distance from village to the centre of district | Far |
|  |  | Medium |
|  |  | Near |
| var7 | Road accessibility of four-wheel motor vehicles within village | Always can be accessed |
|  |  | Can be accessed, except if it rains, erosion, etc |
|  |  | Can be accessed, except during the rainy season |
|  |  | Can not be accessed |

The correlation matrix of those variables at national level is shown below.

Table 2. The Correlation Matrix of Geographic Variables at National Level

|    | var1   | var2   | var3   | var4   | var5   | var6   | var7 |
|----|--------|--------|--------|--------|--------|--------|------|
| v1 | 1      |        |        |        |        |        |      |
| v2 | 0.5137 | 1      |        |        |        |        |      |
| v3 | 0.3233 | 0.2307 | 1      |        |        |        |      |
| v4 | 0.1751 | 0.1134 | 0.1980 | 1      |        |        |      |
| v5 | 0.1420 | 0.0750 | 0.2108 | 0.6214 | 1      |        |      |
| v6 | 0.1600 | 0.1458 | 0.2539 | 0.3225 | 0.2949 | 1      |      |
| v7 | 0.1967 | 0.1140 | 0.2423 | 0.5398 | 0.8023 | 0.3324 | 1    |

From the correlation matrix above, it is known that the correlation coefficient among the variables are always positive so that those variables are appropriate to be processed in the next steps. To construct the composite indicator of geographic difficulty, first of all, the weight of each variables must be determined. Because the geographical variables utilized are in the categorical scale, Polychoric Principal Component Analysis is chosen as a method in weighting process to determine the weight of each variables. Weight of each variables is derived from the first eigen vector produced by Polychoric Principal Component Analysis Process because the first eigen vector explains the largest variability of data. The result of Polychoric Principal Component Analysis Process is shown in Table 3.

Table 3. Weight Resulted From Polychoric Principal Component Analysis

|      | Weight | | | |
|------|--------|--------|--------|--------|
|      | Score1 | Score2 | Score3 | Score4 |
| var1 | -0.756 | -0.385 | -0.215 | 0.142  |
| var2 | -0.515 | -0.231 | 0.133  | .      |
| var3 | -0.707 | -0.369 | 0.131  | .      |
| var4 | -0.953 | -0.613 | -0.328 | 0.217  |
| var5 | -1.069 | -0.657 | 0.121  | .      |
| var6 | -0.395 | -0.002 | 0.392  | .      |
| var7 | -0.803 | -0.545 | -0.455 | 0.160  |

The statistic formulation to construct index of geographic difficulty is

$$I_c = \sum_{a=1}^{7} \sum_{b=1}^{B_a} \gamma_{ab} x_{abc}$$

$I_c$ : geographic difficulty index for the c-th village
$\gamma_{ab}$ : weight for the a-th variable and the b-th score
$x_{abc}$ : dummy variable ($x_{abc} = 1$ if the c-th village has the a-th variable and the b-th score, $x_{abc} = 0$ if the others).

In order to be convenient for interpretation, min-max normalization procedure is implemented so that the geographic difficulty index has the range scale from 0 (low difficulty) to 100 (high difficulty).

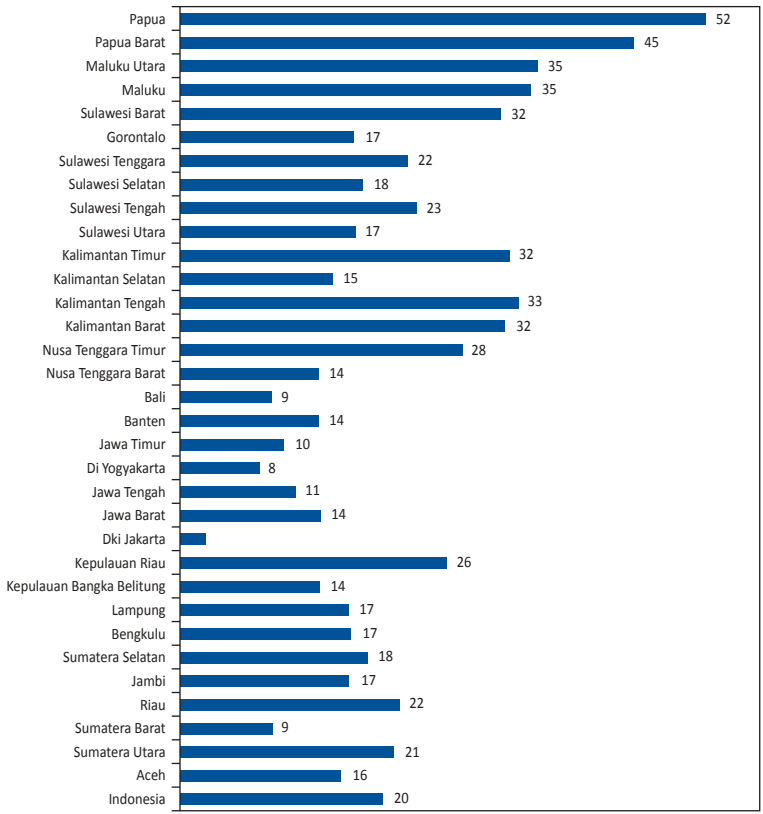

**Figure 1. Average Index Of Geographic Difficulty, Split By Provinces**

The result of geographic difficulty index can be summarized over provinces so that we can perceive the comparison of geographic difficulty level among provinces. As it can be seen in Figure 1, the average geographic difficulty index at national level is 20. The province that has the lowest average index is DKI Jakarta, while Papua has the highest level of difficulty (the average indexes are 2 and 52, respectively). Based on this results, this paper focusses the analysis in Papua Province as the domain estimation of simulation study in terms of stratification method, statistical efficiency, and cost efficiency issues.

### 3. WEALTH INDEX

Population Census 2010 is used as data source to construct wealth index, including 9 operational variables such as level of education of household head, type of floor, main source of lighting, the primary fuel for cooking, the main source of drinking water, type of toilet, septic tank, telephone ownership, and internet ownership. The method used to determine weight is the same with that of Geographic Difficulty Index, namely Polychoric Principal Component Analysis. Firstly, wealth index is calculated in household level. Then, wealth index concentration is constructed in the census block level (census block is usually used as primary sampling unit). The statistic formulation used are as seen below.

$$I_r = \sum_{p=1}^{9} \sum_{q=1}^{Q_p} \gamma_{pq} \, x_{pqr}$$

$I_r$ : wealth index for the r-th household
$\gamma_{pq}$ : weight for the p-th variable and the q-th score
$x_{pqr}$ : dummy variable ($x_{pqr} = 1$ if the r-th household has the p-th variable and the q-th score, $x_{pqr} = 0$ if the others).
Normalization wealth index in household level is calculated by:

$$I_r^{(norm)} = \left(\frac{I_r - \min(I_r)}{\max(I_r) - \min(I_r)}\right) \times 100$$

Then, the wealth concentration index in census block level is calculated by:

$$I_s = \frac{1}{R_s}\sum_{r=1}^{R_s} I_r^{(norm)}$$

Table 4. The Summary of Wealth Index Concentration in Papua Province

| Mean | Std. Dev. | Min | Max | Skewness | Kurtosis |
|---|---|---|---|---|---|
| 38.5756 | 17.9029 | 9.8031 | 88.4714 | 0.6920 | 2.0428 |

**Scenarios for Stratification**

Stratification is constructed based on the wealth consentration index and geographic difficulty index. There are 15 scenarios of stratification simulated which the determination of strata boundaries is based on cumulative root frequency method. Moreover, optimum allocation is implemented so that the sampling variance for population can be calculated to measure the statistical efficiency. On the other hand, in terms of cost efficiency, there is an approximation by using the average of difficulty index of these scenarios. In the other words, it means that if the scenario has the low average of difficulty index, this result indicate that this scenario tends to have the low level of difficulty. As a consequence, it is expected that the level of cost efficiency can be improved if this scenario is implemented.

Based on the simulation study of Papua Province, the correlation coefficient between geographic difficulty index and wealth concentration index is very strong, approximately -0,7983. It means that the higher level of geographic difficulty, the lower level of prosperity. This high correlation has statistical advantage in terms of stratum construction, because one of the most important aspects that influences the efficiency of stratification is the high correlation between variables used in stratification and research variables.

Table below shows the value of variance according to number of strata in which it shows that difference combination of strata resulted in difference value of variances. In the case that strata of wealth index are not used, the higher number of geographic difficulty strata, the lower value of variance will be, but the value of variance increase when the strata of geographic difficulty is set to be 4. Meanwhile, in the case that strata of geographic difficulty are not used, the higher number of wealth index strata, the lower the variance will be. The smallest value of variance will be obtained in the case that each of wealth index strata and geographic difficulty strata are set to be 4.

Table 5. Value of Variance According to Number of Wealth Index Strata and Geographic Difficulty Strata

| Strata of Wealth Index | Strata of Geographic Difficulty | | | |
|---|---|---|---|---|
| | 1 | 2 | 3 | 4 |
| 1 | | 0.074884 | 0.059585 | 0.059765 |
| 2 | 0.059843 | 0.044727 | 0.039180 | 0.039472 |
| 3 | 0.030705 | 0.029911 | 0.028058 | 0.028277 |
| 4 | 0.023989 | 0.023083 | 0.022918 | 0.022916 |

According to simulations, the less difficult (the lowest cost) is strata 13 or by using three strata of geographic difficulty. In addition, based on statistical efficiency, the smallest variances are strata 43 and 44. Furthermore, some sampling scenarios can be categorized into four quadrants, the first quadrant with small sampling variance and low cost, the second quadrant with big sampling variance and low cost, the third quadrant with big sampling variance and high cost, and the fourth quadrant with small sampling variance and high cost. The value of in Papua after using stratification are 37 to 39, compared to 52 in mean of population. It shows that by using stratification, the geographic difficulty is decreasing. According to 15 scenarios used, all values of difficulty are less than population mean, this is the advantage of stratification. The high correlation between wealth index and geographic difficulty index provides advantage to the mean of geographic difficulty index based on the selected samples when using optimum allocation.

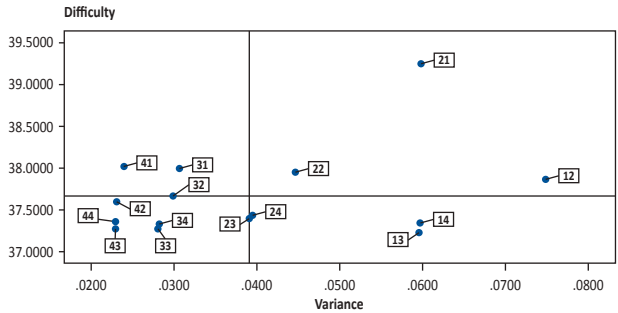

**Figure 2. Sampling Scenarios Based on Variance and Difficulty**

## 5. CONCLUSIONS

The correlation coefficient between geographic difficulty index and wealth concentration index is very strong which means that the higher level of geographic difficulty, the lower level of prosperity. This strong correlation has statistical advantage in terms of stratum construction. Simulation results illustrate that several alternative scenarios of efficient stratification with small sampling variance and low cost of the survey can be obtained. Stratification in remote area impacts on better statistical efficiency and cost efficiency.